\documentclass[twocolumn,8pt, showpacs, aps,  prd,  a4paper, superscriptaddress,  nofootinbib, tightenlines,  floats,floatfix]{revtex4}
\usepackage{graphicx}
\usepackage{epsfig}
\usepackage{subfigure}
\usepackage{latexsym}
\usepackage{amsmath,amssymb}

\newcommand{\beq}{\begin{equation}}
\newcommand{\eeq}{\end{equation}}
\newcommand{\beqa}{\begin{eqnarray}}
\newcommand{\eeqa}{\end{eqnarray}}
\newcommand{\bea}{\begin{array}}
\newcommand{\ena}{\end{array}}

\input colordvi

\def\be{\begin{equation}}
\def\ee{\end{equation}}
\def\bea{\begin{eqnarray}}
\def\eea{\end{eqnarray}}

\def\4pig{\sfrac{4\pi G}{c^{4}}}

\def\hsp5{\hspace{5mm}}
\newcommand{\sfrac}[2]{{\textstyle{#1\over#2}}}
\def\case#1/#2{\textstyle\frac{#1}{#2}}

\begin{document}
\title{Universality of critical magnetic field in holographic superconductor }

\author         {D. Momeni}
\affiliation    {Eurasian International Center for Theoretical Physics and Department of General \& Theoretical Physics, Eurasian National University, \\ Astana 010008, Kazakhstan}

\author         {R. Myrzakulov}
\affiliation    {Eurasian International Center for Theoretical Physics and Department of General \& Theoretical Physics, Eurasian National University, \\ Astana 010008, Kazakhstan}
\date{\today}
 \begin{abstract}
In this letter we study aspects of the holographic superconductors analytically  in the presence of a  constant external magnetic field. We show that the critical temperature and critical magnetic field can be calculated at nonzero temperature. We detect the Meissner effect in such superconductors. A universal relation between black hole mass $ M$ and critical magnetic field $H_c$ is proposed as $\frac{H_c}{M^{2/3}}\leq 0.687365$. We discuss some aspects of phase transition in terms of black hole entropy and the Bekenstein's entropy to energy upper bound.\\
\end{abstract}

\maketitle

 High temperature superconductors are phases of the matter which are quite conductive at temperature close to $T\sim 138^ {o} K$.
It was proven that such superconductors could be described using a gauge / gravity
duality (for a quick review see \cite{Horowitz:2006ct}).  The
gauge/gravity mechanism is based on a one-to-one correspondence
between the exact black hole solutions of a gravitational model on a Riemannian curved manifold of
spacetime with a negative cosmological constant  and a
conformal quantum theory (in flat space-time)\cite{Witten}. In quantum picture,
the quantum objects are relevant quantum operators and the physical
quantities are totally renormalizable.\par
 All these quantum objects live on
a flat spacetime (boundary). 
If we assume that the blackhole horizon is thermal at temperature $T_{BH}$ and the CFT has  an associated nonzero temperature $T_{CFT}$, then it is possible to consider two systems in  thermodynamic equilibrium. It means that $T_{BH}=T_{CFT}$.
This duality backs to the Maldacena's conjecture in string theory as follows:\par Maldacena conjecture
\cite{Maldacena}: Type IIB of superstring theory in bulk with $AdS_5
\times S_5$ geometry is dual, to a specific gauge theory, $N = 4, SU
(N_c) $ super-Yang-Mills theory (SYM) in four dimensions.\par 

Another alternative name for this conjecture  is Anti-de Sitter/conformal field theory
(AdS/CFT) . It is proven that this conjecture plays an important role to describe the
high temperature superconductor (for earlier works see
\cite{gubser}-\cite{HartnollJHEP2008}). Following the basic logic of gauge/gravity, we need to have a one -to-
one correspondence  between each dynamical quantity of the
superconductor (temperature and condensation of the order parameter)
and a set of the parameters for black hole.
Depending on the speciefic type of condensation,the different types of holographic
superconductors were considered by several authors
 \cite{Cai:2014ija}-\cite{Wu:2010ii}. In the high temperature material the superconductivity is a function of the critical value of magnetic field and the subject was studied in several papers
 \cite{Ge:2010aa}-\cite{Roychowdhury:2012hp}.
The Meissner effect was investigated by analytical and numerical methods 
while the external magnetic field was made some important  effects on the basic parameters of
superconductor  \cite{salvio1}-\cite{salvio4}.\par
In this paper we investigate analytically the effect of a constant
external magnetic field on the critical temperature of a type $s$
superconductor. Holographic superconductors with constant external
magnetic field quite were studied  by the other authors. Especially
in \cite{Zeng:2009dv}, a hairy numerical solution was found for any
value of external magnetic field $H$ in range $0 \leq H \leq H_ {max}
$, where the $H_ {max} = 0.687365$.
The significant observation in the value of $H$ was that the
critical temperature $T_c$ in presence of a magnetic field vanishes
when $H$ approaches the critical value $H_c = H_{max}$.\par
 In this
work we revisit analytically the basic properties of such hairy black hole solutions.
 We will treat the equations of motion as a system of
coupled nonlinear differential equations which can be reduced to an
eigenvalue-eigenfunction problem near the criticality $T=T_c$. We
show that it is possible to find semi-exact solutions
for equations  approximately near the critical point. Using these semi-exact solutions we
obtain critical temperature and maximum of external magnetic field.
The appearance of conformal dimensions and uniqueness of the
expectation values of the dual CFT operators are verified. We  observe that for a typical mass of the
black hole ($M$), there is  an upper bound for the external
magnetic field and the size of thermal horizon (in critical point) .
Consequently we show that this maximum size of horizon leads
to a maximum value of the critical temperature.  We compute the
difference of specific heat between normal and superconducting
phases we investigate the stability issue. We prove that
such superconductors remain stable.
In order to make a holographic model for a high temperature
superconductor we need to introduce  an appropriate gravitational
sector that is written in the "GOD given" units as follows:
 \cite{Hart}):
\begin{eqnarray}
\mathcal{L}_g=R-\frac{6}{L^2}-\frac{1}{4}\mathcal{F}_{\mu\nu}\mathcal{F}^{\mu\nu}.
\end{eqnarray}
Where $R$ is the Ricci scalar, $L$ is AdS length, $A_{\mu}$ is gauge
field and $F_{\mu\nu}=\partial_{\mu}A_{\nu}-\partial_{\nu}A_{\mu}$. 
If only one external constant magnetic field $H$ is present, the non
zero components of the $F_{\mu\nu}$ tensor are given by $
F_{xy}=\frac{H}{r^2}$.
We consider an imaginary sphere with radius $r$. We get the total energy density stored in it$
F_{\mu\nu}F^{\mu\nu}\sim H^2.
$
The gravitational field solution is a
 magnetized Schwarzschild Anti-de Sitter (SAdS) black hole in planar coordinates $ (t, r, x, y) $\cite{L. J. Romans} :
\begin{eqnarray}
&&ds^2=-f(r)dt^2+\frac{dr^2}{f(r)}+r^2(dx^2+dy^2)\label{g},\\&&\nonumber f(r)   \equiv f=r^2-\frac{M}{r}+\frac{H^2}{r^2}.
\end{eqnarray}
The black hole has a horizon which is real solution of the equation $f(r_{+})=0$. The function $f\equiv f(r)$ can be rewritten as follows:
\begin{eqnarray}
f=r^2-\frac{r_{+}^3}{r}-\frac{H^2}{rr_{+}}+\frac{H^2}{r^2}.
\end{eqnarray}
The temperature of the
black hole is defined by:
\begin{eqnarray}
T=\frac{f'(r_{+})}{4\pi}=\frac{r_{+}(3-h^2)}{4\pi}
\end{eqnarray}
 \par
We define the dimensionless magnetic field  as  $h^2=\frac{H^2}{r_{+} ^4} $ and now we have
built fully gravitational sector. For matter sector we use a
Lagrangian similar to Ginzburg-Landau (GL) which it includs a Maxwell
field $A=A_{\mu}$ and a charged complex scalar field $\psi$
\cite{Gubser:2008px,Hartnoll:2008vx}:
\begin{equation}
{\cal L}_m=-\frac{1}{q}\Big(\frac{1}{4}F^{ab}F_{ab}+m^2|\psi|^2+|\nabla \psi-iA\psi|^2\Big)\label{Lm}.
\end{equation}
Note that the (\ref{Lm}) is different from the original  GL theory,
 firstly, the mass term is negative  $m_{\psi}^2<0$ secondly, there is no potential term  $|\psi|^4$ here. For stabilization we need an
AdS bulk geometry.  Our guess is that a physical mechanism like
 Higgs mechanism launched in the region outside of the horizon\cite{Gubser:2008px}. Due to the symmetry of the metric and gauge freedom we set
  $A_t=\phi(r)$ , $\psi^{*}(r)=\psi(r)$. So,the equations of motion may be recast to the following forms :
\begin{eqnarray}
g^{\mu\nu}(\nabla_{\mu}-i A_{\mu})(\nabla_{\nu}-i A_{\nu})\psi-m^2\psi=0\label{kg}.\\
\nabla_{\mu}F^{\mu\nu}=i(\psi^{*}\nabla^{\nu}\psi-\psi\nabla^{\nu}\psi^{*})+|\psi|^2A^{\nu}\label{maxwell}.
\end{eqnarray}
 The equations of motion are written as follows in case of metric (\ref{g}) :
\begin{eqnarray}\label{eqn:2ode}
&&\psi''+(\frac{f'}{f}+\frac{2}{r})\psi'+\frac{\phi^2}{f^2}\psi-\frac{m^2}{f}\psi=0,\nonumber\\
&&\phi''+\frac{2}{r}\phi'-\frac{2\psi^2}{f}\phi=0
\end{eqnarray}
We put $L=1$ and note that the mass of a scalar field is $m^2 >m_{BF}^2=-\frac{9}{4}$.\par
It is convenient to rewrite the equations in terms of dimensionless coordinate $z=\frac{r_{+}} {r} $
\begin{eqnarray}
&&\psi''(z)+\frac{f'}{f}\psi'(z)+\frac{r_{+}^2}{z^4}\Big(\frac{\phi(z)^2}{f(z)^2}-\frac{m^2}{f(z)}\Big)\psi(z)=0,\label{psi(z)}\\&&
\phi''(z)+\frac{m^2r_{+}^2\psi(z)^2}{z^4f(z)}\phi(z)=0\label{phi(z)},
\end{eqnarray}
Now the metric function is
 $f\equiv f (z) =r_ {+} ^2 (z^ {-2} -h^2z-z+h^2z^2) $. \par
We would like to find  the solutions of
nonlinear equations 
 (\ref{psi(z)},\ref{phi(z)}) . For $\{\phi,\psi\}$, Eqs.  (\ref{psi(z)},\ref{phi(z)}) can be solved as:
\begin{eqnarray}
&&\phi(z)=\phi^{0}(z)-m^2r_{+}^2\int_{0}^{z}{\frac{(z-w)\psi^2 (w)\phi(w)dw}{w^4f(w)}}\\&&
\psi(z)=\psi^{+}(z)+\psi^{-}(z)\\&&\nonumber+\Big(-\psi^{+}(z)\int{\frac{\psi^{-}(z)p(z)}{W(\psi^{+}(z),\psi^{-}(z))dz}}\\&&\nonumber+\psi^{-}(z)\int{\frac{\psi^{+}(z)p(z)}{W(\psi^{+}(z),\psi^{-}(z))dz}}\Big).
\end{eqnarray}
In which :
\begin{eqnarray}
&&\frac{d^2}{dz^2}\{\phi^{0}(z)\}=0,
\{\frac{d^2}{dz^2}+\frac{f'}{f}\frac{d}{dz}-\frac{m^2r_{+}^2}{z^4f(z)}\}\psi^{\pm}(z)=0,
\\&&
p(z)=-\frac{r_{+}^2\psi(z)}{z^4}\frac{\phi(z)^2}{f(z)^2},\\&&
W(\psi^{+},\psi^{-})=\psi^{+}(\psi^{-})'-\psi^{-}(\psi^{+})'.
\end{eqnarray}
The AdS/CFT mechanism is focused on the behavior of the fields at the AdS boundary, i.e.
 $z=0$, when  $f\sim r_ {+} ^2z^ {-2} $, we are able to approximate solutions to (\ref{psi(z)},\ref{phi(z)}) :
\begin{eqnarray}
\phi^{0}(z)=c_0+c_1 z,\ \
\psi^{\pm}(z)=c_{\pm}z^{\Delta_{\pm}}.
\end{eqnarray}
here the conformal dimension is $\Delta_{\pm}=\frac{3}{2}\pm\sqrt{m^2+\frac{9}{4}}$ . At this precise expression, $c_{\pm}= \frac{<\mathcal{O}_{\pm}>}{r_{+}^{\Delta_{\pm}}}$. We conclude that precise form of the Wronskian is:
\begin{eqnarray}
W(\psi^{+}(z),\psi^{-}(z))=-\Big(2c_{+}c_{-}\sqrt{m^2+\frac{9}{4}} \Big)z^{2}.
\end{eqnarray}
We need to estimate the following pair of integrals:
\begin{eqnarray}
I_1=\int_{0}^{z}{\frac{(z-w)\psi^2 (w)\phi(w)dw}{w^4f(w)},
I_{\pm}=\int{\frac{\psi^{\pm}(z)p(z)dz}{W(\psi^{+}(z),\psi^{-}(z))}}}.
\end{eqnarray}
following the above information we know that:
$$
p(z)|_{z=0}\sim -r_{+}^{-2}c_{\pm}z^{\Delta_{\pm}}c_{0}^2.
$$
Thus we have:
\begin{eqnarray}
I_{\pm}=A_{\pm}\times z^{2\pm2\sqrt{m^2+\frac{9}{4}}}.
\end{eqnarray}
The most common technique for obtaining precise information about the form of solutions within an interval $z\in[0,1]$ is to evaluate   $\psi_{z=0}$:
\begin{eqnarray}
&&\psi=c_{+}z^{\Delta_{+}}+c_{-}z^{\Delta_{-}}+\Big(c_{-}A_{+}\times z^{\frac{7}{2}+\sqrt{m^2+\frac{9}{4}}}\\&&\nonumber-c_{+}A_{-}\times z^{\frac{7}{2}-\sqrt{m^2+\frac{9}{4}}}\Big).
\end{eqnarray}
May integral $I_1$ can be adequately approximated by an additive integration whose output is a sum over several smaller dimensional expressions:
\begin{eqnarray}
&&I_1=\int^{z}{\frac{(z-w)\psi^2 (w)\phi(w)dw}{w^4f(w)}}=\\&&\nonumber z\{\frac{\psi^2 (w)\phi(w)}{w^4f(w)}\}|_{z=\epsilon}-\{w\frac{\psi^2 (w)\phi(w)}{w^4f(w)}\}_{w=\epsilon},\ \ \epsilon\sim0.
\end{eqnarray}
We placed the three expressions individually and was finally able to write $I_1$:
\begin{eqnarray}
I_1=\alpha_0 z+\beta_0.
\end{eqnarray}
In which
$\alpha_0=\frac{(c_{\pm})^2c_0\epsilon^{2\Delta_{\pm}-2}}{r_{+}^2},
\beta_0=\frac{(c_{\pm})^2c_0\epsilon^{2\Delta_{\pm}-1}}{r_{+}^2}$.
By collecting results from the past, the solution gives shape to  the forms of fields in vicinity
of AdS boundary :
$$\phi(z)=\mu-\frac{\rho}{r_{+}} z ,
\ \ 
\psi(z)=\frac{<\mathcal{O}_{+}>}{r_{+}^{\Delta_{+}}}z^{\Delta_{+}}
$$

Moreover, our attempts to establish the form of solutions in the vicinity of the AdS boundary was also fruitless.

The quantization of the relevant operators has been investigated and in some cases a very good agreement with CFT results was found $\mathcal{O}_{-}=0\Rightarrow c_ {-} =0$.
Here $\mathcal{O}_{+} $ is an operator of dimension $\Delta_{+} $ in dual field theory and $<\mathcal{O}_{+} >$ corresponds to the vacuum expectation value of it. We will introduce a set of the boundary conditions
\cite{Gubser:2008px,Hartnoll:2008vx}:
\begin{eqnarray}
\phi(1)=0,\ \ (h^2-3)\psi'(1)=m^2\psi(1).
\end{eqnarray}
\par
Best advice is to use  continuously and to always expand the fields somewhere in the AdS boundary and horizon
 \cite{Gregory}. 
Conventional computing method only provide facilities for exact matching, a more adaptive approach is required in order to support numerical aggregation.
Both fields smoothly across a surface with location $z=z_m\in[0,1]$
\cite{Ge:2011cw}-\cite{Roychowdhury:2012hp}. 
We conclude that precise evaluation of expectation values of relevant operators  is required for proper integration of actin into counters:
\begin{eqnarray}
\frac{<\mathcal{O}_{\pm}>}{r_{+}^{\Delta_{\pm}}}=\frac{1}{2\pi i}   \oint_{\gamma} \frac{\psi(t)dt}{t^{\Delta_{\pm}+1}}
\end{eqnarray}
The vacuum expectation values of relevant operators are shown on the boundary, but no bulk term.
At finite temperature we have $h\neq\sqrt{3}$. 
The analytical solutions in the AdS sector, in the vicinity of $z=0$, computed temporally 
as follows:
\begin{eqnarray}
&&\phi(z)=a\Big[(1-z)-\frac{m^2r_{+} b^2}{8\pi T}(1-z)^2\Big] ,\\
&&\psi(z)=b\Big[1-\frac{m^2}{h^2-3}(1-z)\\&&\nonumber+\frac{(1-z)^2}{2}\Big(-\frac{a^2}{32\pi^2T^2}+\frac{3m^2(1-h^2)}{(h^2-3)^2}\Big)\Big]
\end{eqnarray}
here $a=-\phi'(1), b=\psi(1)$. 
The process by which one obtains coefficents from equations of motion in the presence of boundaries $z=0,z=1§$ is called matching method:

\begin{eqnarray}
&&a ( \frac{1}{2}-\frac{1}{32}\,{\frac {{m}^{2}r_{+}{b}^{2}}{\pi\,T}})
=\mu- \,{\frac {\rho}{2r_{+}}}\label{eq1},
\\
&&b \Big( 1-\frac{1}{2}\frac {m^{2}}{{h}^{2}-3}-\frac {1}{256}\frac
{a^2}{\pi ^{2}{T}^{2}}+\frac{3}{8}\frac {m^2 ( 1-{h}^{2}
 )  ^{2}}{ ( {h}^{2}-3) ^{2}) }\Big)\\&&\nonumber=
\frac {<\mathcal{O}_{+}>(\frac{1}{2}) ^{\Delta}}{{r_{+}}^{\Delta}}\label{eq2}.\\
&&a \Big( -1+\frac{1}{8}\frac {{m}^{2}r_{+}{b}^{2}}{\pi\,T}\Big)
=-\frac { \rho}{r_{+}}\label{eq3},
\\
&&b \Big( \frac {m^2}{h^{2}-3}+\frac {1}{64}\frac {a^{2}
}{{\pi }^{2}{T}^{2}}-\frac{3}{2}\frac {m^2 (1-{h}^{2})
  ^{2}}{ ({h}^{2}-3 ) ^{2}} \Big) =\\&&\nonumber\frac {2\Delta<\mathcal{O}_{+}>
 ( \frac{1}{2}) ^{\Delta}}{{r_{+}}^{\Delta}}\label{eq4}.
\end{eqnarray}
For the comparison of analytical with numerical result \cite{wen}, relative  value $m^2=-2$ were used.
This system (\ref{eq1},\ref{eq3})  gives us:
\begin{eqnarray}
&&a=4\mu-\frac{3\rho}{r_{+}},\ \ b^2=
\frac{8\pi T}{m^2}\Big[\frac{1}r_{+}-\frac{\rho}{ar_{+}^2}\Big]\\&&\nonumber\Rightarrow ar_{+c}=\rho \Leftrightarrow <\mathcal{O}_{+}>|_{r_{+}\rightarrow r_{+c}}=0
\end{eqnarray}
According to the definition of dual quantities, we know that:
$$
 r_{+c}=\frac{\rho}{\mu}\Rightarrow r_{+c}^4-Mr_{+c}+H_c^2=0.
$$
In critical mode, the definition of critical temperature  is given, followed by the abbreviation in parentheses:
\begin{eqnarray}
T_c=\frac{3r_{+c}^4-H_c^2}{4\pi r_{+c}^3}\label{Tc}
\end{eqnarray}
Equally, it is clear that the elimination of  $r_ {+c} $ in $T_c$ will lead to :
\begin{eqnarray}
&& 256\,{T_c}^{4}{\pi }^{4}{H_c}^{4}+96\,{T_c}^{2}{\pi }^{2}{M}^{2}{H_c}^{2}-27\,
{M}^{4}\\&&\nonumber+64\,{T_c}^{3}{\pi }^{3}{M}^{3}+256\,{H_c}^{6}=0\label{secular-eq}.
\end{eqnarray}
$T_c (H) $ and $H_c$ in (\ref{Tc}) or (\ref{secular-eq}) are defined
as a critical point in the presence of external magnetic field and
critical magnetic field that turns the superconducting state into a
normal state below the $T_ {c0} = \frac{4\pi r_ {+c0}} {3} $ (with
no magnetic field). The maximum of magnetic field $H_ {Max} $ is
obtained at the minimum critical temperature $T_c (H) = 0$, hence to
obtain the maximum magnetic field $H_ {Max} $, we simply let the
temperature in (\ref{secular-eq}) vanish and obtain:
\begin{eqnarray}
\frac{H_{Max}}{M^{2/3}} = \sqrt[6]{\frac{27}{256}} = 0.687365.
\end{eqnarray}
which is exactly the numerical result of \cite{Zeng:2009dv}.\par
We analyze the pair $ ( H_c, T_c)$ carefully.
First when there is no magnetic field $H=0$, horizon size tends towards zero $r_{+}=0$.
It is also worth remembering that there was  famous the entropy of black hole which became simple in our case:
$$S=\frac{\pi r_{+}^2}{l_{p}^2}
$$
We know $S\geq 0$. 
At normal phase, where there is no entropy , this time external magnetic field disappears altogether.
Thus from $S$ we see that the entropy change of a system during for a normal process is zero. An example of its success is in determining the normal phase of a holographic superconductor. Things tend to move toward disorder we think its called second order phase transition from normal to superconducting phase. For enough large external magnetic field, equilibration occurs initially at large values of $H_c$ and proceeds away from the Bekenstein's entropy bound $\frac{S}{E}<2\pi R$ \cite{Bekenstein:1980jp}.
We showed how gravitational entropy can be defined in general for holographic superconductor.
Does the critical system tend toward the maximum of entropy production and maximum $H_c$?. The answer is yes. If it becomes more disordered(superconducting), its entropy and $H_c$ increase. External magnetic field $H_c$ maximum property derived from Bekenstein's entropy bound principle.
We listed $H_{Max},r_{c Max}$ for different values of  masses  in the following Table:
\begin{center}
    \begin{tabular}{| l || l | l | l |}
    \hline
    Mass (M)  & 0.7 & 0.5 & 0.3 \\ \hline
   $H_{Max}$ & $\simeq 0.3$ & $\simeq 0.4 $ & $\simeq 0.5$ \\ \hline
    $r_{c}^{Max}$ & $\in[0.6,0.7]$ & $\in[0.7,0.8]$ &$\in[0.8,0.9]$\\
    \hline
    \end{tabular}
\end{center}
\par
The adjacency $ (H_c, T_c) $   of this magnetized superconductor initially located at the top of the $T_c$ then describes the result of the critical magnetization. This behavior clearly displays higher levels of $H_c$ in holographic rather than  the mass of the black
hole in each case. This analysis shows how the mass $M$ builds to a critical magnetic field $H_c$, then rapidly falls. No direct stress is applied against the initial external magnetic field $H_c$. This result provides data on the occurrence of  critical magnetic field $H_c$   in holographic superconductor from $0$ to $H_c^ {Max}$.
You may increase or decrease the the critical magnetic field $H_c$  simply by   decreasing  or extending the the
critical temperature $T_c$, or by increasing or decreasing the amount of starter $H_c$ to the $H_c^ {Max}$.
The results obtained were consistent and showed a strong dependence on the  the critical magnetic field $H_c$.
The results of $H_c$  using analytical method were consistent with the observations of the figure II of
\cite{Nakano-Wen,wen}. 
With the increase in $M$ and $H_c$ the amount
of $T^ {Max} $ increases.
 
The following table indicates the
variability of $ (M, H_c^ {Max}, T_c^ {Max}) $.

\begin{center}
    \begin{tabular}{| l || l | l | l |}
    \hline
    Mass (M)  & 0.3 & 0.5 & 0.7 \\ \hline
   $H_{Max}$ & $ 0.3$ & $\simeq 0.42 $ & $\simeq 0.55$ \\ \hline
    $T_{c}^{Max}$  & $\simeq 0.16$ & $\simeq 0.18$ & $\simeq 0.21$ \\
    \hline
    \end{tabular}
\end{center}

To complete the discussion of the thermodynamic point of view, we
refer to the Rutgers formula. This formula indicates the
discontinuity in specific heat and it shows the difference between
specific heat of pure metals at the critical temperature in the
superconducting and normal phase\cite{book1,book2}:
\begin{eqnarray}
&&\Delta C=C_{\text{superconducting}}-C_{normal}
\\&&\nonumber=\mu_0 T\Big(H_c(T)\frac{d^2 H_c(T)}{dT^2}+\Big[\frac{dH_c(T)}{dT}\Big]^2\Big).\label{DeltaC}
\end{eqnarray}
At the critical temperature $T_ {c0} $ that is the critical point in the absence of external magnetic field, and $H_c (T_ {c0}) = 0$, therefore, the first term in the Rutgers formula (\ref{DeltaC}) vanishes and becomes the form of :
\begin{eqnarray}
\Delta C=\mu_0 T\Big[\frac{dH_c(T)}{dT}\Big]^2.\label{DeltaC2}
\end{eqnarray}
From the Rutgers formula (\ref{DeltaC2}) at the critical point, one can apparently observe that, $\Delta C$ must be positive.
It is adequate to study the system after an infinitesimal deviation
from the Universal bound, when $H_c=M^{2/3}u-\delta H_c$ we obtain:
\begin{eqnarray}
&&\delta(\Delta C)={\Delta C}|_{H_c=M^{2/3}u}-{\Delta C}|_{H_c=M^{2/3}u-\delta H_c}>0.
\end{eqnarray}
So the system is stable under this bound.
We note that  the phase transition was studied here is  type I to type II. The reason is that 
under the magnetic field  $H_c$ there is Meissner effect. When
the magnetic field $H$ becomes stronger, we will observe the quantized
vortices. 
The results show  a rapid increase in $H_ {c}$ in the superconducting phase and the more $H $ increase through the the normal phase system.
 Finally we
note that the case of $T=0$ which is equal to $h=\sqrt{3}$ in
reference \cite{wen} were studied and here is not checked.
Holographic superconductors are gravitational
analogy of high temperature superconductors in condensed matter
physics. Fundamental properties of this class of  superconductors can be simulated
to the properties of a black hole in the gravitational sector in
this model. In this paper we reviewed the effect of a constant external magnetic
field on a superconductor type s were. We showed that
 the matching method  can give a critical
magnetic field and critical temperature in non-zero temperature. The study of the behavior
of the critical temperature and critical magnetic field has shown
that with increasing critical magnetic field the critical
temperature decreases and approaches to zero at a maximum magnetic
field. We have also shown that the superconducting state is
consistent with the existence of a thermodynamic state with maximal
entropy. Universal relation between the magnetic field for $H_c$ and
the mass of black hole $M$ is as follow $\frac{H_c}{M^{2/3}}\leq
\sqrt[6]{\frac{27}{256}} = 0.687365$.

We thank the referee(s) for useful comments.


\begin{thebibliography}{90}


\bibitem{Horowitz:2006ct}
Horowitz   G~T~and ~Polchinski J, 2006
  In *Oriti, D. (ed.): Approaches to quantum gravity* 169-186
  [gr-qc/0602037]

\bibitem{Witten}
Witten E, 1998 Adv. Theor. Math. Phys. {\bf 2}, 253 

\bibitem{Maldacena}
 Maldacena J, 1999 Adv. Theor. Math. Phys. {\bf 2}, 231[
Int. J. Theor. Phys. {\bf 38}, 1113 (1999)]

\bibitem{gubser}
Gubser S S,2008  Phys. Rev. D {\bf 78}, 065034 

\bibitem{Hart}
 Hartnoll S A , Herzog C P , and  Horowitz G T ,2008  Phys. Rev. Lett. {\bf
101}, 031601 

\bibitem{HartnollRev}
Hartnoll S A,2009 Class. Quant. Grav. {\bf 26}, 224002 

\bibitem{HerzogRev}
Herzog C P,2009 J. Phys. A {\bf 42}, 343001 

\bibitem{HorowitzRev}
 Horowitz G T,2010 arXiv: 1002.1722 [hep-th]


\bibitem{HartnollJHEP2008}
 Hartnoll S A,  Herzog C P and Horowitz G T ,2008 J. High Energy
Phys. {\bf 12}, 015  [arXiv:0810.1563]

\bibitem{Cai:2014ija}
 Cai R~-G~, ~Li L, Li L~-F~ and Yang R~-Q~,2014
  JHEP {\bf 1404}, 016 
  [arXiv:1401.3974 [gr-qc]]

\bibitem{Cai:2013aca}
Cai  R~-G~, Li L~ and Li L~-F~,2014
  JHEP {\bf 1401}, 032 
  [arXiv:1309.4877 [hep-th]]
\bibitem{Momeni:2012ab}
 Momeni D~, Majd N~ and Myrzakulov R~,2012
  Europhys.\ Lett.\  {\bf 97}, 61001 
  [arXiv:1204.1246 [hep-th]]
\bibitem{Roychowdhury:2013aua}
  Roychowdhury D~,2013
  JHEP {\bf 1305}, 162 
  [arXiv:1304.6171 [hep-th]]
\bibitem{Cai:2013oma}
Cai  R~-G~, Li L~, Li L~-F~ and Su R~-K~,2013
  JHEP {\bf 1306}  063
  [arXiv:1303.4828 [hep-th]]
\bibitem{Arias:2012py}
 Arias R~E~ and Landea I~S~,2013 
  JHEP {\bf 1301} 157
  [arXiv:1210.6823 [hep-th]]
\bibitem{Gangopadhyay:2012gx}
  Gangopadhyay S~and Roychowdhury D~,2012
  JHEP {\bf 1208} 104
  [arXiv:1207.5605 [hep-th]]
\bibitem{Chen:2012kc}
Chen  S~, Pan Q~ and Jing J~, 2013
  Commun.\  Theor.\  Phys.\  60, {\bf 471}
  [arXiv:1206.5462 [gr-qc]]
\bibitem{Cai:2012nm}
Cai  R~-G~, He S~, Li L~and Zhang Y~-L~,2012 
  JHEP {\bf 1207} 027
  [arXiv:1204.5962 [hep-th]]
\bibitem{Kuang:2011dy}
  Kuang X~-M~, Li W~-J~and Ling Y~,2012
  Class.\ Quant.\ Grav.\  {\bf 29}  085015
  [arXiv:1106.0784 [hep-th]]
\bibitem{Murray:2011gr}
Murray  J~M~ and Tesanovic Z~,2011 
  Phys.\ Rev.\ D {\bf 83} 126011
  [arXiv:1103.3232 [hep-th]]
\bibitem{Cai:2010zm}
  Cai R~-G~, Nie Z~-Y~and Zhang H~-Q~,2011
  Phys.\ Rev.\ D {\bf 83}  066013
  [arXiv:1012.5559 [hep-th]]
\bibitem{Momeni:2013fma}
Momeni  D~, Myrzakulov R~and Raza M~,2013
  Int.\ J.\ Mod.\ Phys.\ A {\bf 28}, 1350096 
  [arXiv:1307.8348 [hep-th]]
\bibitem{Zeng:2010fs}
 Zeng H~-B~, Sun W~-M~ and Zong H~-S~,2011
  Phys.\ Rev.\ D {\bf 83}, 046010 
  [arXiv:1010.5039 [hep-th]]
\bibitem{Cai:2010cv}
Cai  R~-G~, Nie Z~-Y~and Zhang H~-Q~,2010
  Phys.\ Rev.\ D {\bf 82}, 066007 
  [arXiv:1007.3321 [hep-th]]
\bibitem{Zeng:2009dr}
Zeng  H~-B~, Fan Z~-Y~and Zong H~-S~, 2010
  Phys.\ Rev.\ D {\bf 81}, 106001
  [arXiv:0912.4928 [hep-th]]
\bibitem{Chen:2010mk}
Chen  J~-W~, Kao Y~-J~, Maity D~, Wen W~-Y~ and Yeh C~-P~, 2010
  Phys.\ Rev.\ D {\bf 81} 106008
  [arXiv:1003.2991 [hep-th]]
\bibitem{Zeng:2010vp}
  Zeng H~-B~, Fan Z~-Y~ and Zong H~-S~,2010
  Phys.\ Rev.\ D {\bf 82} 126008
  [arXiv:1007.4151 [hep-th]]

\bibitem{cpl10}. Chen S-B,Pa Q-Y,Jing J-L,2012 Chinese Physics B,,21(4):040403
\bibitem{Gao:2011aa}
Gao  D~,2012
  Phys.\ Lett.\ A {\bf 376}  1705
  [arXiv:1112.2422 [hep-th]]
\bibitem{Ge:2012vp}
 Ge X~-H~, Tu S~F~ and Wang B~,2012
  JHEP {\bf 1209}  088
  [arXiv:1209.4272 [hep-th]]
\bibitem{Krikun:2013iha}
 Krikun A~,2013
  arXiv:1312.1588 [hep-th]
\bibitem{Nishida:2014lta}
Nishida  M~,2014
  arXiv:1403.6070 [hep-th]
\bibitem{Li:2014wca}
 Li L~-F~, Cai R~-G~, Li L~and Wang Y~-Q~,2014
  arXiv:1405.0382 [hep-th]
\bibitem{Momeni:2013via}
Momeni  D~, Raza M~and Myrzakulov R~,2014
  Eur.\ Phys.\ J.\ Plus {\bf 129}, 30 
  [arXiv:1307.2497 [hep-th]]
\bibitem{Momeni:2013waa}
 Momeni D~, Raza M~, Setare M~R~and Myrzakulov R~,2013
  Int.\ J.\ Theor.\ Phys.\  {\bf 52}, 2773 
  [arXiv:1305.5163 [physics.gen-ph]]
\bibitem{Momeni:2013eva}
Momeni  D~, Raza M~and Myrzakulov R~, 2013
  J.\ Grav.\  {\bf 2013}, 782512
  [arXiv:1305.3541 [physics.gen-ph]]
\bibitem{Momeni:2012tw}
Momeni  D~, Myrzakulov R~, Sebastiani L~and Setare M~R~,2012
  arXiv:1210.7965 [hep-th]
\bibitem{Momeni:2012uc}
Momeni  D~, Setare M~R~ and Myrzakulov R~, 2012
  Int.\ J.\ Mod.\ Phys.\ A {\bf 27}, 1250128
  [arXiv:1209.3104 [physics.gen-ph]]
\bibitem{Setare:2011ip}
Setare   M~R~and Momeni D~,2011
  Europhys.\ Lett.\  {\bf 96}, 60006 
  [arXiv:1106.1025 [physics.gen-ph]]
\bibitem{Momeni:2011ca}
  Momeni D~and Setare M~R~,2011
  Mod.\ Phys.\ Lett.\ A {\bf 26}, 2889 
  [arXiv:1106.0431 [physics.gen-ph]]
\bibitem{Momeni:2010jf}
Momeni D~,   Setare M~R~ and Majd N~,2011
  JHEP {\bf 1105}, 118 
  [arXiv:1003.0376 [hep-th]]
\bibitem{Cai:2011tm}
Cai  R~-G~, Li L~, Zhang H~-Q~and Zhang Y~-L~,2011
  Phys.\ Rev.\ D {\bf 84}, 126008 
  [arXiv:1109.5885 [hep-th]]
\bibitem{Wu:2010ii}
Wu  J~-P~,2010
  arXiv:1006.0456 [hep-th]
\bibitem{Ge:2010aa}
  Ge X~-H~, Wang B~, Wu S~-F~ and G~-H~Yang,2010
  JHEP {\bf 1008}, 108 
  [arXiv:1002.4901 [hep-th]]
\bibitem{salvio1}
Domenech O , Montull M , Pomarol A ,  Salvio A, Silva P J ,2010 JHEP 1008:033,,arXiv:1005.1776.
 \bibitem{salvio2}Montull M~, Pujolas O~,
 Salvio A~ and Silva P~J~, 2011
  Phys.\ Rev.\ Lett.\  {\bf 107}, 181601
  [arXiv:1105.5392 [hep-th]]
\bibitem{salvio3}  Montull M~, Pujolas O~,
 Salvio A~ and Silva P~J~,2012
  JHEP {\bf 1204}, 135 
  [arXiv:1202.0006 [hep-th]]
\bibitem{salvio4}  Salvio A~, 2012
  JHEP {\bf 1209}, 134
  [arXiv:1207.3800 [hep-th]]
\bibitem{Zeng:2009dv}
Zeng H ,  Fan Z,  Ren Z,2009
  Phys.\ Rev.\ D {\bf 80}, 066001 
  [arXiv:0906.2323 [hep-th]]


\bibitem{L. J. Romans}
Romans L J ,  1992 Nucl. Phys. B 383, 395 [arXiv:hep-th/9203018]

\bibitem{Gubser:2008px}
Gubser  S~S~,2008
 ,  arXiv:0801.2977 [hep-th]


\bibitem{Hartnoll:2008vx}
Hartnoll  S~A~, Herzog C~P~ and Horowitz G~T~,2008
  arXiv:0803.3295 [hep-th]
 
\bibitem{Yin:2013fwa}
Yin  L~, Hou D~ and Ren H.~-c~,2013
  arXiv:1311.3847 [hep-th]
\bibitem{Gregory}
 Gregory R, Kanno S , and Soda J, 2009 J. High Energy Phys. {\bf 10},
010 
\bibitem{Ge:2011cw}
Ge  X~-H~ and Leng H~-Q~,2012
  Prog.\ Theor.\ Phys.\  {\bf 128}, 1211
  [arXiv:1105.4333 [hep-th]]
\bibitem{Cui:2013uha}
Cui  S~-l~ and Xue Z~,2013
  arXiv:1306.2013 [hep-th]
\bibitem{Zhao:2013pva}
 Zhao Z~, Pan Q~ and Jing J~,2013
  arXiv:1311.6260 [hep-th]
\bibitem{Roychowdhury:2012hp}
  Roychowdhury D~,2012
  Phys.\ Rev.\ D {\bf 86}, 106009 
  [arXiv:1211.0904 [hep-th]]

\bibitem{Bekenstein:1980jp}
Bekenstein  J~D~, 1981
  Phys.\ Rev.\ D {\bf 23}, 287
\bibitem{Nakano-Wen}
Nakano E, Wen W Y , 2008 Phys. Rev. D {\bf 78}, 046004 
\bibitem{wen}
Momeni D ,  Nakano E, Setare M R ,  Wen W-Y, 2013 Int. J.  Mod. Phys. A
 28,  8  1350024,[arXiv:1108.4340]
\bibitem{book1}
Ehrenfest P,1934 Comm. Phys. Lab. Leiden,suppl.75b
\bibitem{book2}
 Rutgers  N, 1934 Physica 1,1055




\end{thebibliography}
\end{document}